\documentclass{elsart}

 \usepackage{epsfig}

\usepackage{amssymb}

\begin{document}

\begin{frontmatter}





\title{A Model for the  Pion Structure Function}


\author{F. Bissey, J. R.  Cudell, J. Cugnon, M. Jaminon,}
\author{J. P. Lansberg and P. Stassart}

\address{Universit\'e de  Li\`ege, D\'epartement de  Physique B5, Sart Tilman,
\\ B-4000 LIEGE 1, Belgium}

\begin{abstract}
The pion structure function is investigated in a simple model, where pion and
constituent quark fields are coupled through the simplest pseudoscalar
coupling. The imaginary part of the forward $\gamma^{\star} \pi \rightarrow
\gamma^{\star}\pi$ scattering amplitude is evaluated and related to the
structure functions. It is shown that the introduction of non-perturbative
effects, linked to the size of the pion and preserving gauge invariance, allows
a connection with the quark distribution. It is predicted that
higher-twist terms become negligible for $Q^2$ larger than $\sim$2~GeV$^2$ and
that quarks in the pion have a momentum fraction smaller than in the proton
case.
\end{abstract}

\end{frontmatter}


\section{Introduction}

Deep-inelastic scattering (DIS) experiments provide us with a wealth
of information about the structure of hadrons, usually cast in the
form of structure functions. These data are only partly understood
in the framework of quantum chromodynamics (QCD). Indeed, perturbative QCD is
consistent with the \( Q^{2} \)
evolution of the structure functions at
sufficiently high \( Q^{2} \) and $x$~\cite{COO97}. However, it is neither able
to predict the structure functions themselves, as the latter are supposed
to result from non-perturbative effects, among which confinement and
spontaneous chiral symmetry breaking, nor the magnitude of the initial value
$Q^{2}_0$ from which the  $Q^{2}$ dependence can be evaluated .
The interest has recently  widened to off-diagonal parton
distributions~\cite{JI97,RA96,CO97,VA98},
which potentially offer to reach complementary information, especially
about parton correlations.

Phenomenological quark models, which possess some non-perturbative
aspects and which are rather successful in reproducing low-energy
properties of hadrons, are expected to  help us to
understand the connection between DIS data and non-pertubative
aspects. Pions and other low-mass
mesons are the simplest systems for which effective models exist that
incorporate, in some simplified
way, such special QCD features such as spontaneous chiral symmetry breaking and
anomalies.

Our original plan was to investigate the properties of off-diagonal parton
distributions in a simple model for light mesons. However, we realized that the
theoretical investigation of the diagonal distributions using phenomenological
quark models is far from being settled.
There have been several theoretical investigations along these lines
in recent
years~\cite{JA80,SH93,DA95,WE99}. 
They rely on the assumption that distributions evaluated in 
leading-twist approximation at small \( Q^{2} \), where these models 
apply,  can serve as input in the DGLAP evolution 
equations~\cite{AL77} to generate parton distributions that are directly
comparable with experimental data at 
large \( Q^{2} \)~\cite{OW84,AU89,GL92,GRS,SU92}.
This is a
rather tricky point as it is not clear that a good approximation at low $Q^{2}$
 can be evolved by perturbative equations to  large $Q^{2}$, as it is not
sure that, in this regime, the forward $\gamma^{\star} \pi \rightarrow
\gamma^{\star} \pi$ amplitude
can be parametrised in terms of parton distributions (despite the
existence of low-$Q^2$ parametrisations \cite{GRS,SU92}).

Some of the theoretical works mentioned above have been carried out
in the framework of the Nambu and Jona-Lasinio (NJL) model~\cite{NA61}. This
model  is indeed quite successful in reproducing low-energy phenomenology and
embodies chiral symmetry breaking, which is  believed to be crucial for this
success.  However,  investigations of the structure functions 
have given rather different
results~\cite{SH93,DA95,WE99}. This situation originates from the fact that the
NJL model
needs to be regularized and that different regularizations yield different
results. 

We show in this paper that these complications may in fact be avoided.  We
investigate the simplest model of a
pion, in which the $q \overline{q} \pi$ vertex is represented by the simplest
pseudoscalar coupling. Of course, the quark
fields and the coupling constant should reflect in some way the properties
of actual pions. One possible guide is provided by the large \( N_{c} \)
limit of the NJL model. The latter is equivalent, in the chirally
broken phase, to a \( \sigma  \)-model with massive constituent quarks
and an effective pion-quark-quark coupling constant. Our starting point
for the investigation of the parton distributions of the
pion consists in a Lagrangian that includes massive pion and massive quark
fields interacting through the simplest pseudoscalar vertex.
We show below that there is no need to regularize this model as the imaginary
part of the
forward $\gamma^{\star} \pi \rightarrow \gamma^{\star} \pi$ scattering
amplitude is finite. However, the interaction
between the constituent quarks should vanish when their relative momentum
is large enough. It is possible to cope with this requirement
by imposing a finite momentum cut-off, which mimics the effect of a pion wave
function. Such a  step leads to
the appearance of a straightforward relation between the \( \gamma ^{\star
}-\pi  \)
cross sections and the quark distributions at high enough $Q^2$,
as we shall show below.

\begin{figure}
\centering {\epsfig{file=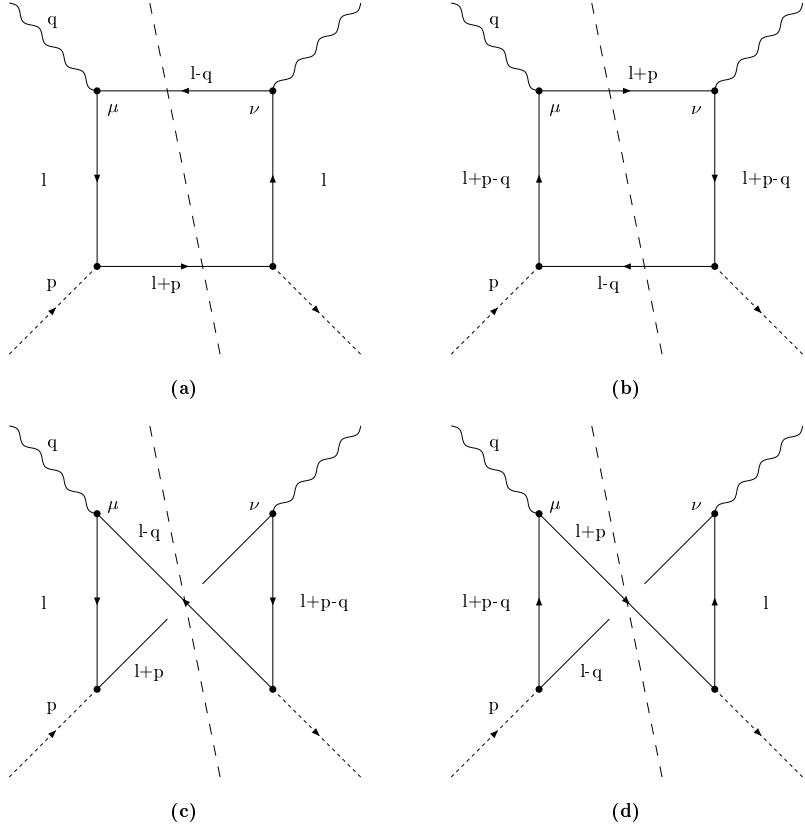,width=10cm}}
\caption{Simplest diagrams contributing to the imaginary part of the forward
amplitude for the scattering of a virtual photon by a neutral pion.  Upper
(lower) diagrams are referred to  as box (crossed) diagrams. Dashed lines
represent the discontinuity of the amplitudes or their imaginary parts.}
\label{f1}
\end{figure}

\section{The model}

We consider an isospin
triplet pion field $\vec{\pi}=(\pi^+,\pi^0,\pi^-)$ interacting with quark
fields $\psi$ through the  Lagrangian density
\begin{equation}
 {\mathcal L}_{int}= i g (\overline{\psi}\mbox{  $\vec{\tau}$} \gamma_5 \psi) .
\mbox{  $\vec{\pi}$}, \label{Lint}
\end{equation}
where $\vec{\tau}$ is the isospin vector operator. It is our purpose  to
calculate the imaginary part of the forward elastic $\gamma^{\star}-\pi$
scattering amplitude, or equivalently the total cross-section, in the simplest
approximation and extract from it the pion structure functions $W_1$ and $W_2$.
Finally, we want to see whether they are reducible to quark distributions.
We give below explicit results for the neutral pion.
\par
The relevant diagrams contributing to the imaginary part of the forward elastic
scattering  amplitude, up to first order in $\alpha$, the fine structure
constant, and to second order in $g$, are given in Fig.~\ref{f1}. We define the
pion 4-momentum as $p$, the photon 4-momentum as $q$, the ($u$ and $d$) quark
constituent mass as $m_q$ and we use $p^2=m_{\pi}^2$, $q^2=-Q^2$, $p.q=\nu$,
$x=Q^2/(2 \nu)$. This  leads to $s=m_{\pi}^2+Q^2(1/x-1)$. The imaginary part of
the amplitudes can be written, using Cutkosky rules, as
\begin{equation}
\mbox{Im}T^a_{\mu \nu}=  -2 C g^2\int d^4l\{ t^1_{\mu \nu} D_1 D_4 D_1 D_2 \} ,
\label{Ta}
\end{equation}
\begin{equation}
\mbox{Im}T^c_{\mu \nu}=  -2 C g^2\int d^4l\{ t^2 _{\mu \nu}D_1 D_4 D_3 D_2 \}, \label{Tc}
\end{equation}
with the fermionic traces
\begin{eqnarray}
t^1_{\mu \nu}&=&\mbox{Tr}\left[ \gamma_\mu (\gamma.(l-q)+m_q) \gamma_\nu \right.
\nonumber\\
& &\left. (\gamma.l+m_q) \gamma_5
(\gamma.(l+p)+m_q) \gamma_5 (\gamma.l+m_q)\right] , \label{t1}\\
t^2_{\mu \nu}&=&\mbox{Tr}\left[ \gamma_\mu (\gamma.(l-q)+m_q) \gamma_5
(\gamma.(l+p-q)+m_q) \gamma_\nu \right.\nonumber\\
& &\left. (\gamma.(l+p)+m_q)
 \gamma_5 (\gamma.l+m_q)\right], \label{t2}
\end{eqnarray}
and the fermion propagators (or their contribution to the cuts)
\begin{equation}
\begin{array}{rclrcl}
D_1^{-1} & = & l^2-m_q^2, & \quad D_2  & = & 2 \pi 
\delta\left((l+p)^2-m_q^2\right), \\
D_3^{-1} & = & (l+p-q)^2-m_q^2, & \quad D_4 & = & 2 \pi 
\delta\left((l-q)^2-m_q^2\right).
\end{array}\label{prop}
\end{equation}

The constant $C=5 e^2/(384 \pi^6)$ accounts for  flavour, charge and loop
momentum integration factors in  the particular case (neutral pion) under
consideration. Diagrams (b) and (d) have the same contributions as (a) and (c),
respectively. The second ones are obtained from the first ones  by the 
substitution $l \rightarrow q-p-l$ in the integrals. Using $T^1=T^a+T^b$,
$T^2=T^c+T^d$ and $T=T^1+T^2$, one can rewrite
\begin{equation}
\mbox{Im}T^1_{\mu \nu}=  -2 C g^2\int d^4l\{ t^1_{\mu \nu} 
D_1 D_4 D_1 D_2 + (l
\rightarrow q-p-l)\} , \label{T1}
\end{equation}
\begin{equation}
\mbox{Im}T^2_{\mu \nu}=  -2 C g^2\int d^4l\{ t^2_{\mu \nu} D_1 D_4 D_3 D_2 + (l
\rightarrow q-p-l)\}, \label{T2}
\end{equation}
which makes gauge invariance ($q^{\mu} \mbox{Im} T_{\mu \nu}
=\mbox{Im} T_{\mu \nu} q^{\nu}=0$)  explicit by  inspection of the integrand.
It is worth  emphasizing that the sum of the four diagrams is gauge
invariant, but none of them alone has this property.

We perform the integrations  in Eqs.~\ref{T1} and \ref{T2} using  Sudakov
variables. We define $l=-\xi p + \eta q + l_t$, with  $l_t$ perpendicular
to both $p$ and $q$. A little algebra shows that
\begin{equation}
d^4l \quad \!\!\!\delta((l+p)^2-m_q^2) \delta((l-q)^2-m_q^2)
=\frac{dl^2 d\phi }{4 \sqrt{\nu^2+2\nu m_{\pi}^2
x}}, \label{jac}
\end{equation}
with the change of variable
\begin{equation}
\xi = x +
{  (l^2-m^2-2m_\pi^2)(2 x -1)
\over
2\nu + 4 m_\pi^2 x},
\end{equation}
and with $\eta$ and $l_{t}^2$ related to $l^2$ through the relations
\begin{eqnarray}
\eta&=&
{ - l^2 + m^2 + m_\pi^2 (2 x -1)\over
2 \nu + 4 m_\pi^2 x}
+ {m_\pi^2 ( - l^2 + m^2)\over
\nu(2\nu + 4 m_\pi^2 x)},\\
l_{t}^2&=& (l^2-m_\pi^2 x)(1-x) + m^2 x\nonumber\\
&+& {(1-x) ( -4 m_\pi^2 x (l^2-m_\pi^2 x)
+ (l^2 - m^2)^2)\over
2 \nu + 4 m_\pi^2 x}\nonumber\\
&+&{m_\pi^2( l^2-4 m^2 x(x-1)
 - m^2 - m_\pi^2 x) \over
2 \nu + 4 m_\pi^2 x}\nonumber\\
&+& {m_\pi^2 (l^2- m^2)^2\over
4 \nu^2 + 8 \nu m_\pi^2 x}.
\end{eqnarray}

The integration over the azimuthal angle $\phi$ is trivial and one is left only
with the integration over the variable $\tau \equiv -l^2$.
The integration bounds on $\tau$ come from the positivity of the energies of
the  on-shell intermediate states (see Fig.~\ref{f1}) and from the space-like
definition of $l_t$ (yielding  $l_{t0}^2 \leq 0$), which introduces the most
stringent constraints. The latter take the following forms:

\begin{eqnarray}
\tau \leq 2 \nu  &+& \frac{((1-x) x m_{\pi}^2-(2-x)m_q^2)}{1-x} \nonumber\\
&-& \frac{(m_q^2-x^2 m_{\pi}^2)(m_q^2-(1-x)^2 m_{\pi}^2)}{2 \nu(1-x)^2 },
 \label{con3}
\end{eqnarray}
and
\begin{equation}
\tau \geq    \frac{(m_q^2-(1-x)m_{\pi}^2)}{1-x}
  + \frac{(m_q^2-x^2 m_{\pi}^2)(m_q^2-(1-x)^2 m_{\pi}^2)}{2 \nu(1-x)^2 },
\label{con4}
\end{equation}
where only the first terms in the expansion of the r.h.s. in $1/\nu$ are given
(the full expressions are used in numerical evaluations).

The structure functions $W_1$ and $W_2$ being defined by
\begin{eqnarray}
W_{\mu \nu}&=&\frac{1}{2\pi} \mbox{Im}T_{\mu \nu}\nonumber\\
&=& \left(-g_{\mu
\nu}+\frac{q_{\mu}q_{\nu}}{q^2}\right) W_1 +
\left(p_{\mu}-q_{\mu}\frac{p.q}{q^2}\right)
\left(p_{\nu}-q_{\nu}\frac{p.q}{q^2}\right)W_2, \label{W}
\end{eqnarray}
 they can be calculated through the contractions
\begin{equation}
W_{\mu}^{\mu}=\frac{-6x W_1 +(\nu + 2 m_{\pi}^2 x)W_2}{2x} , \label{Wc1}
\end{equation}
\begin{equation}
p^{\mu}p^{\nu}W_{\mu \nu}=\frac{-2x (\nu + 2 m_{\pi}^2  x) W_1 +(\nu + 2
m_{\pi}^2 x)^2 W_2}{4 x^2}. \label{Wc2}
\end{equation}
We also checked that we get the same results for $W_1$ and $W_2$ by computing
directly the  $\gamma^{\star} \pi \rightarrow q \overline{q}$ transverse and
longitudinal cross-sections.
\par
Although we give below results for the exact expressions, it is worth giving
the large-$Q^2$ limit:
\begin{equation}
W_1=\frac{5 g^2}{24 \pi^2} \left[ \ln\left(\frac{2  (1-x) \nu}{M^2}\right) -
\frac{m_{\pi}^2} {M^2} x (1-x)  \right], \label{Lnu}
\end{equation}
with $M^2=m_q^2-m_{\pi}^2 x (1-x)$.

At this  point, the pion cannot be interpreted as a collection of partons with
a probability distribution. Indeed, the crossed diagrams are not suppressed by
a power of $Q^2$   compared to the box diagrams and  therefore do not allow
such an interpretation. This is manifest when their respective contributions
are written down in the combined small-$m_{\pi}$, large-$Q^2$ limit:
\begin{equation}
W_1^{box}=\frac{5 g^2}{24 \pi^2} \left[ \ln\left(\frac{2  (1-x)
\nu}{m_q^2}\right)-1\right], \hspace{0.5cm} W_1^{cross}=\frac{5 g^2}{24 \pi^2}.
\label{cross}
\end{equation}
A reason for the appearance of this undesirable feature is that we have not yet
imposed the fact that the pion has a finite size. The simplest way to do this
is to require that the square of the relative four-momentum of the quarks
inside the pion is limited to a maximum value $\Lambda^2$. The former  quantity
is given by
\begin{equation}
O_1=(p + 2l)^2= -2\tau + 2 m_q^2 - m_{\pi}^2,\label{O1}
\end{equation}
for vertices like those in diagram Fig.~1.(a), and by
\begin{equation}
O_2=(p + 2(l+p-q))^2= 2\tau + 6 m_q^2 - m_{\pi}^2 -4 \nu,\label{O2}
\end{equation}
for vertices like those in diagram  Fig.~1.(b). We either require $|O_1| <
\Lambda^2$ or  $|O_2| < \Lambda^2$ for all diagrams, which excludes the
interval $[\Lambda^2, 2 \nu-\Lambda^2/2]$ from the $\tau$ integral. 

$|O_1|$ and $|O_2|$ cannot be small simultaneously. The crossed diagrams
have their main contribution for $O_1\simeq O_2$, and are thus suppressed
by a power $\Lambda^2/Q^2$ when the cut is imposed. The box diagrams
have a leading contribution for $|O_1|$ or $|O_2|$ small, and are not
power suppressed by the cut.
 
\begin{figure}
\centering {\epsfig{file=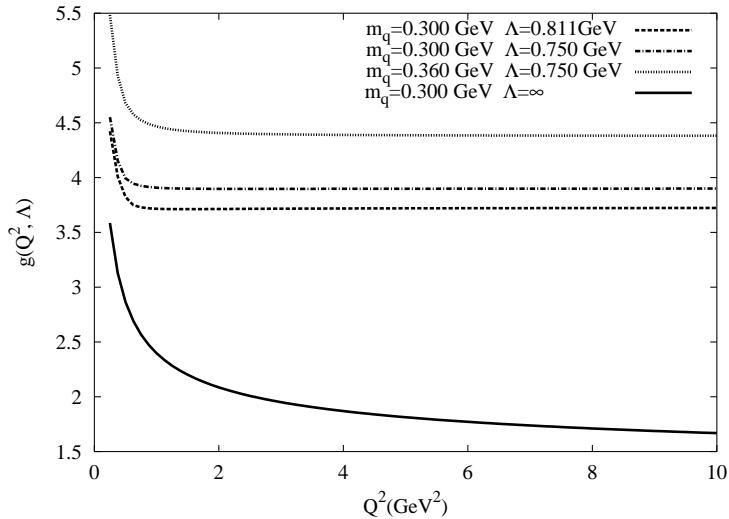,width=10cm}}
\caption{Values of the coupling constant $g(Q^2,\Lambda)$ which fulfill  sum
rule (Eq.~\ref{normF1}), for the values of the parameters indicated at the
top.}
\label{f2}
\end{figure}

Physically,
this happens because, for the crossed diagrams, the  momentum transfer suffered
 by the quark (antiquark) has to be re-emitted by the antiquark (quark), which
is impossible when this momentum transfer becomes too large. For the box
diagrams, the transferred momentum  is taken and released by the same quark (or
antiquark) and there is no suppression. Note that $\tau=-t$ (for the
$\gamma^{\star}\pi \rightarrow q \overline{q}$ process) is a physical
observable quantity. This guarantees that our implementation of the cut-off  is
gauge-invariant.  With our cut-off, the crossed diagrams appear now as higher
twists: they are suppressed  at least as $1/Q^2$ and the box diagram
contribution  can now be interpreted in terms of parton distributions.
Indeed, we checked numerically that, for typical values of $\Lambda$ (see
Fig.~\ref{f2}), the ratio between the respective magnitudes of the crossed and box
diagrams is roughly $\sim$0.06$/Q^2$~GeV$^{-2}$, for $x$=0.1; without cut-off,
this ratio is a sizable fraction of unity for  $Q^2$ less than $\sim$ 1000
GeV$^2$.
\begin{figure}
\centering {\epsfig{file=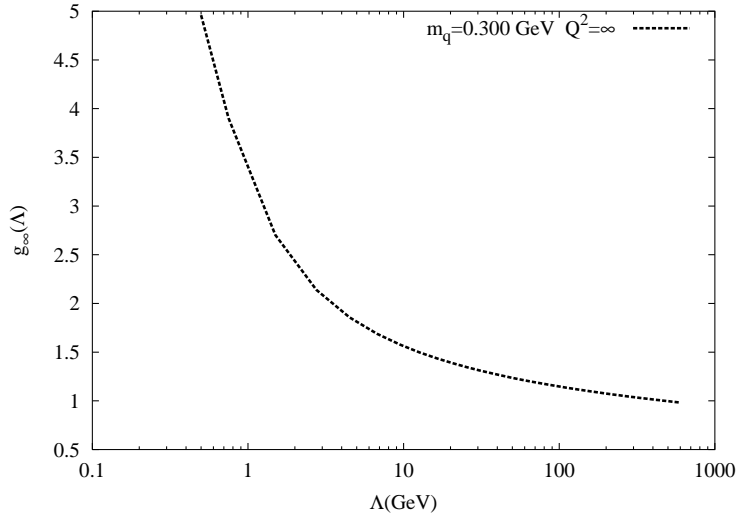,width=10cm}}
\caption{Asymptotic value (for large $Q^2$) of the coupling constant fulfilling
 sum rule (Eq.~\ref{normF1}), as a function of the cut-off parameter
$\Lambda$.}
\label{f3}
\end{figure}

\section{Results}

The structure functions can now be related to the (valence) quark distributions:
\begin{equation}
F_1= W_1=\frac{4}{18}(u_v(x)+\bar{u}_v(x)) + \frac{1}{18}(d_v(x)+\bar{d}_v(x)),
\label{F1}
\end{equation}
\begin{equation}
F_2=\nu  W_2=2x F_1.
\label{F2}
\end{equation}
We stress that the last relation, known as the Callan-Gross relation, comes out
of our calculation. This does indicate that our approximations have been done
consistently\footnote{For charged pions, the additional diagrams implying a
direct coupling of  the virtual photon to the pion are suppressed by a factor
$1/s$ and the leading-twist results are the same, except for the charge
coefficients entering Eq.~\ref{F1}.}.
\par
So far, we have only described the model. In order to make predictions, we need
to fix its parameters, namely \( \Lambda  \), \( m_{q} \) and \( g \). The
latter can be thought of as the normalisation of the quark wave function, and
we determine it by imposing that there are only two constituent quarks in the
pion. In our model, the valence quark distributions are equal
\begin{equation}
u_v(x)=\bar{u}_v(x)=d_v(x)=\bar{d}_v(x)\equiv v(x).
\label{v}
\end{equation}
The condition \( \int _{0}^{1}v(x)dx=1/2 \) then gives us
\begin{equation}
\label{normF1}
\int _{0}^{1}F_{1}(x)dx=\frac{5}{18}.
\end{equation}
 As \( F_{1} \) is a function, not only of \( m_{q} \) and \( \Lambda  \),
but also of \( Q^{2} \), this gives us a coupling constant that evolves with
\( Q^{2} \). The resulting values of \( g \) are shown in Fig. \ref{f2}: for a
finite
cut-off \( \Lambda  \), the cross-section at fixed \( g \) would grow with
energy, until the pion reaches its maximum allowed size, in which case the
cross
section would remain constant. If we impose relation (\ref{normF1}),
this means that \( g(Q^{2}) \) will first decrease until the cut-off makes
it reach a plateau value for \( Q^{2}\gg \Lambda ^{2} \) (in practice, the
plateau value is reached around \( Q^{2} \approx 2\Lambda ^{2} \)). The plateau
value
depends on \( m_{q}/\Lambda  \) (and \( m_{\pi }/\Lambda  \)) and is shown
in Fig. \ref{f3}.

To constrain further our parameters, we can try to use the momentum sum rule
\begin{equation}
\label{mf}
2\left<x\right>=4\int _{0}^{1}xv(x)dx=\frac{18}{5}\int _{0}^{1}F_{2}(x)dx.
\end{equation}
In the parton model, this integral should be equal to 1 as \( Q^{2}\rightarrow
\infty  \),
as we do not have gluons in the model.
\begin{figure}
\centering {\epsfig{file=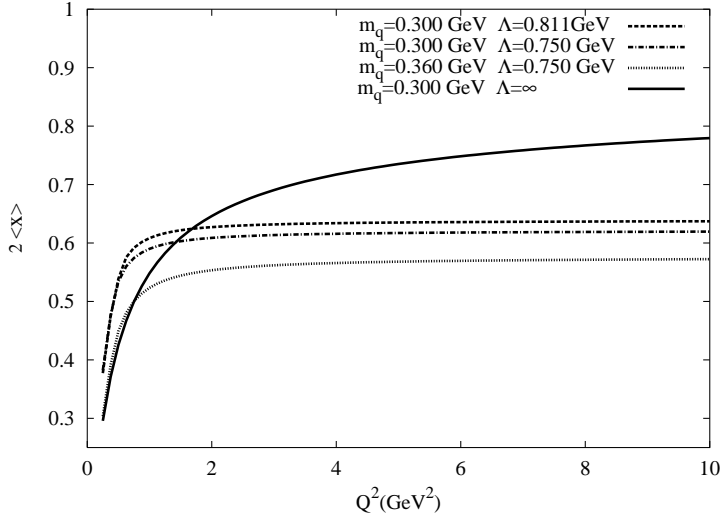,width=10cm}}
\renewcommand{\baselinestretch}{0.5}
\caption{Momentum fraction of the quarks inside the neutral pion
(Eq.~\ref{mf}), as a function of  $Q^2$, for the values of the parameters
indicated at the top of the figure.}
\label{f4}
\end{figure}
\begin{figure}
\centering {\epsfig{file=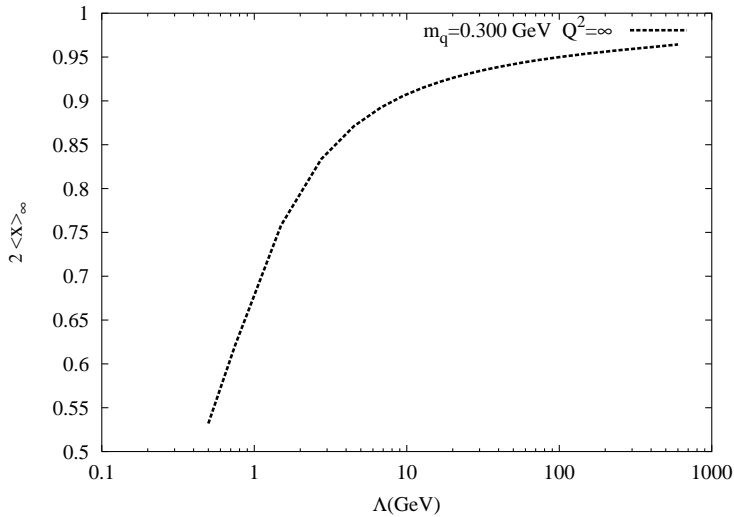,width=10cm}}
\renewcommand{\baselinestretch}{0.5}
\caption{Asymptotic value (for large $Q^2$) of the momentum fraction of the
quarks inside the neutral pion (Eq.~\ref{mf}), as a function of the cut-off
$\Lambda$.}
\label{f5}
\end{figure}
Indeed, there are two limits of
our model that fulfill this condition automatically. First of all, if
we do not impose that the pion has a finite size, then asymptotically we get
\( 2\left<x\right>= \)\( 1 \). In the limit \( m_{\pi }=0 \), we obtain
\begin{equation}
\label{mfch}
2\left<x\right>(\Lambda =\infty , m_{\pi }=0)=\frac{4\ln (2\nu /m_{q}^{2})-3}{4\ln (2\nu
/m_{q}^{2})-1}.
\end{equation}
This means that in that regime, for sufficiently high \( Q^{2} \), the quarks
behave as free particles, and the usual derivation based on the OPE holds
\cite{IT}. The
second case where this holds confirms this interpretation: if we impose 
\break \(
m_{q}\leq m_{\pi }/2 \),
the expressions we have given develop an infrared divergence, which corresponds
to the case where both quarks emerging from the pion are on-shell and free.
This divergence can be re-absorbed into the normalisation (\ref{normF1})
of \( g \), and the sum rule \( 2\left<x\right>=1 \) is again automatically satisfied at
large $Q^2$. 

However,
in the physical pion case, it makes more sense to consider that one of the
quarks remains
off-shell: the imposition of a cut-off changes the sum rule value,
as the fields can never be considered as free. 
Hence, because of the Goldstone
nature
of the pion, one expects that the momentum sum rule will take a smaller value
than in the case of other hadrons. This may explain why fits that assume
the same momentum fraction for valence quarks in protons and pions \cite{GRS}
do not seem to leave any room for sea quarks \cite{Zeus}.
The results for \( 2\left<x\right> \) are
shown in Fig. \ref{f4}. Again, the curves show a plateau at sufficiently large
\( Q^{2} \),
with a value depending on \( m_{q}/\Lambda  \) and \( m_{\pi }/\Lambda  \).
It is easy to show that this value is always smaller than 1. It is shown in
Fig. \ref{f5} as a function of \( \Lambda  \).

To fix the remaining parameters, we must then use our knowledge of constituent
quarks, and of the pion. We choose conservative values \( m_{q}=300 \)~MeV
or \( 360 \)~MeV, and a pion radius of the order of \( 0.25 \) fm, which
corresponds
to \( \Lambda \simeq 800 \)~MeV.
This choice of parameters gives us a momentum fraction \( 2\left<x\right> \) between 0.55
and 0.65 (see Fig. 4), and corresponds to a coupling \( g \) with the plateau
value of 3.8. It is remarkable that this value is very close to the one that
guarantees in the NJL model, with the same constituent mass and the same
cut-off, a unit value for the residue of the $q\overline{q}$ propagator at the
pion pole and the correct value of the electric form factor at
$Q^2=0$~\cite{njll}. In other words, this corresponds to the value  needed for
the pion to appear in the NJL model as constituted of a quark and an 
antiquark\footnote{One should note that, although the numerical values are compatible,
the cut-off has different physical origins in the two approaches: it
corresponds to the maximum internal momentum of the pion in our approach, while
it is a parameter for regularizing ultraviolet divergences in the NJL model.}.
\begin{figure}
\begin{minipage}[t]{6.5cm}
{\epsfig{file=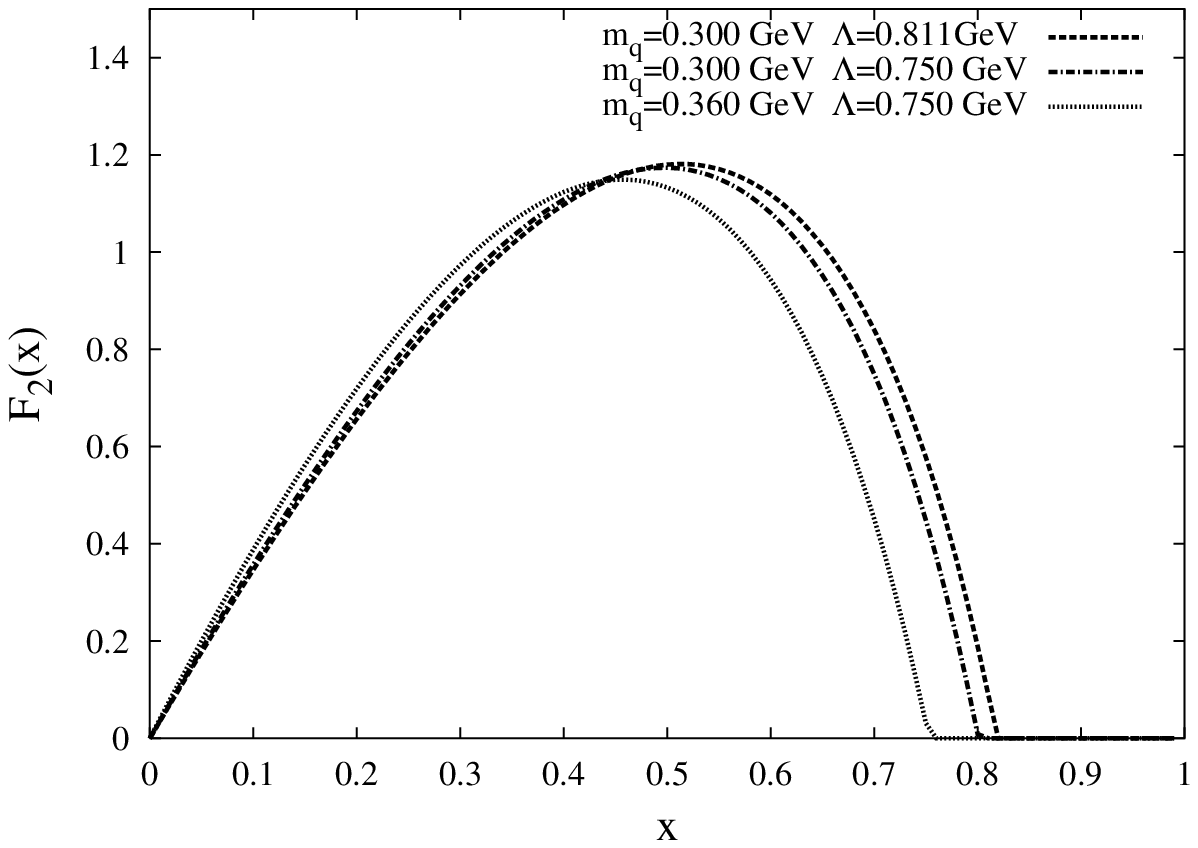,width=6.5cm}}

\end{minipage}
\hfill
\begin{minipage}[t]{6.5cm}
{\epsfig{file=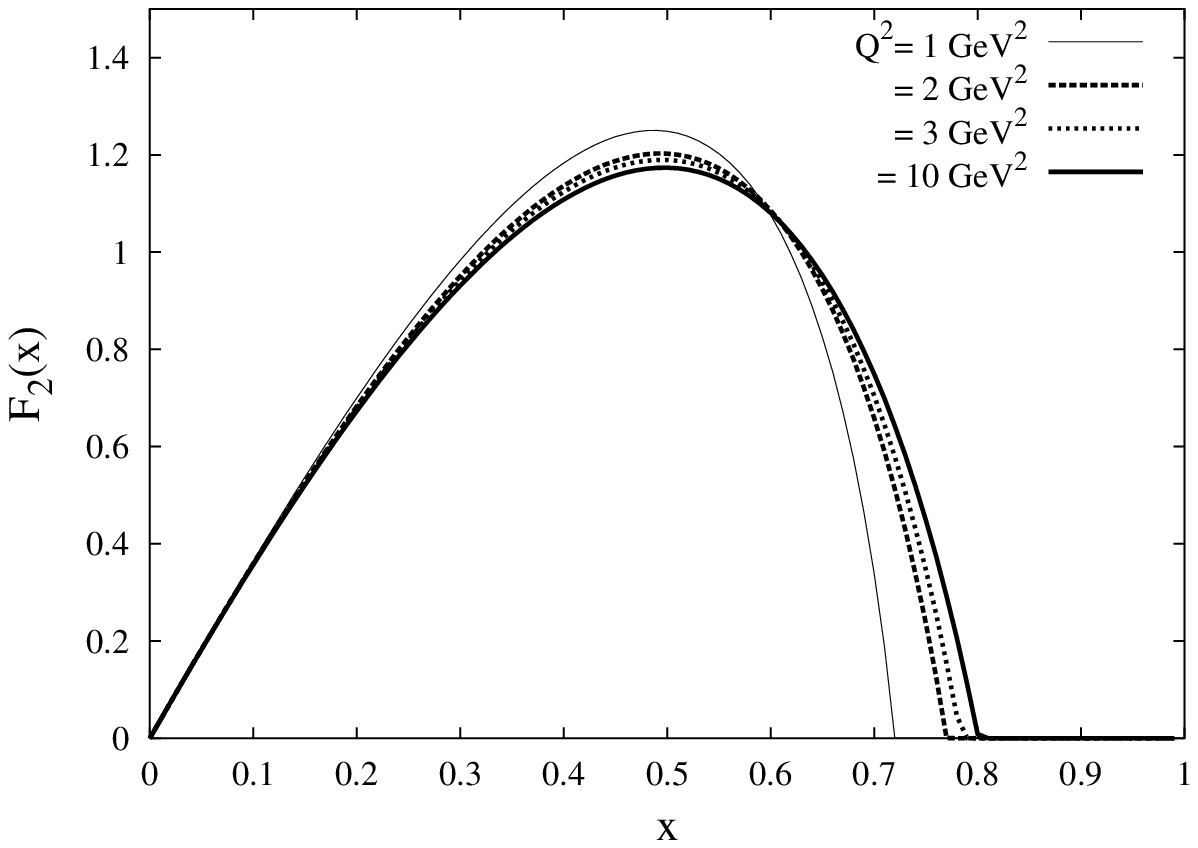,width=6.5cm}}

\end{minipage}
\caption{Structure function $F_2$ for the neutral pion. On the left,
$Q^2$=2~GeV$^2$ and the values of the parameters $m_q$ and $\Lambda$ are as
indicated. On the right, $m_q$=0.3~GeV and  $\Lambda$=0.75~GeV and the values
of $Q^2$ are as indicated.}
\label{f6}

\end{figure}
\par Let us finally examine the properties of the distribution $v(x)$,
or equivalently, of the function $F_2$. Some of our results are summarized in
Fig.~\ref{f6}, for  \break $Q^2$ =  2~GeV$^2$. The most striking feature is the
vanishing of this function for $x$ larger than some value $x_{max}$. This is
again due to the fact  that, because the  quarks are not free in this model,
the actual value of their mass does matter. It can  clearly be seen that this
effect  originates from kinematical cuts: in the chiral limit and for large
$\nu$, the condition $s \geq 4 m_q^2$ is equivalent to
$x_{max} \approx 1-\frac{2 m_q^2}{\nu},$
 in the absence of cut-off.

In the case of a finite cut-off ($\Lambda^2 \ll Q^2$), conditions
(\ref{con3}-\ref{con4}) lower the value of $x_{max}$ to
\begin{equation}
x_{max} \approx 1-\frac{m_q^2}{\Lambda^2}.
\end{equation}
This  means that, for a finite $Q^2$, when $x$ is large, which corresponds to a
small value of $s$, there is no way to put the cut quarks on their mass shell:
this requires at least an energy of 4  $m_q^2$. For given  $Q^2$ and
$\Lambda$, the available energy is increasing with decreasing $x$. As a result,
putting the cut quarks on-shell will be easier for small  than for large $x$.
Therefore, the $x$-distribution ($F_1$) is expected to be enhanced on the low
$x$ side, leading to a momentum fraction smaller than unity.  The vanishing at
large $x$ is not obtained in similar works, in particular in the one of Ref.
\cite{SH93}. In this reference, the Bjorken limit is taken first and the
kinematical constraint (Eq.~\ref{con3}) is not applied . This procedure is
certainly not correct for evaluating  cross-sections at finite $Q^2$. Except
for this vanishing, our results basically agree with those of Ref. \cite{SH93}.
Therefore we will not present results for the DGLAP evolution of our structure
function $F_2$, which we expect to be compatible with the existing data.

\section{Discussion and conclusion}
We have discussed the simplest model allowing to relate  virtual photon-pion
forward elastic scattering to quark distributions. In this model, the
imaginary part of the forward elastic scattering does not show any divergence.
However, this quantity cannot provide the quark distributions  readily
since the numerical importance of the so-called crossed diagrams precludes the
existence of such a relationship. The introduction of a cut-off for the
relative momentum of the quarks inside the pion allows such an interpretation:
crossed diagrams then appear  as  higher twists. The introduction of the
cut-off does not allow to fulfill the momentum sum rule (2$\left<x\right>=1$)
at infinite $Q^2$ because, in the case of the pion, 
constituent quarks can never be considered as free.
We have mentioned above  how  the kinematical constraint, allowing quarks
to be put on their mass shell, leads to the reduction of the momentum fraction.
 This can  easily be seen from our results for the chiral limit (Eq.
\ref{mfch}).

We motivated the cut-off as a manifestation of the pion size. 
The value \break $\Lambda^{-1}=(0.75$~GeV$)^{-1}$ is close to the 
hard core rms radius of the chiral bag model, 0.35 fm \cite{BR86}.

The cut-off has been imposed on the relative quark momentum. This procedure is
at variance with the double-subtraction Pauli-Villars procedure proposed in
Ref.~\cite{WE99}. We have considered such a procedure in our model, but it
produces discontinuities in the structure function as well as a  negative value
for this quantity in some regions of $x$. This procedure is also different from
with the other ones introduced in similar works based on NJL models, where they
are part of  the necessary regularization of these models to avoid divergences.
In addition, there is no need in our approach to consider additional diagrams
with local pion-pion-quark-quark interactions. Yet, the pion-quark-quark
coupling constant turns out to be the same in our 
approach and in NJL models. This
may not be too surprising, as in both cases this coupling is determined by the
requirement that the pion appears as made of two constituent quarks.

\par
Our main conclusion is that pions are  different from other hadrons, in the
sense that the quark momentum fraction should be smaller, and that higher twist
terms disappear for $Q^2 \sim$ 2~GeV$^2$.

\vspace{1cm}

This work has been performed in the frame of the ESOP collaboration (European
Union contract N$^{\circ}$ HPRN-CT-2000-00130). We thank Dr. M. Diehl and  Dr.
P. Guichon for their useful comments.



\end{document}